\documentclass[]{elsart}
\usepackage{longtable}
\usepackage{epsfig}
\usepackage{amssymb}
\usepackage{amsmath}
\usepackage{bm}
\usepackage{graphicx}
\usepackage{graphics}
\usepackage{color}
\usepackage{pstricks}
\usepackage{cite}

\newcommand{\rr}{{\mathbf r}}
\newcommand{\PhiPhi}{\bm \Phi}

\newcommand{\vv}{{\mathbf v}}
\newcommand{\ww}{{\mathbf w}}
\newcommand{\uu}{{\mathbf u}}

\newcommand{\nn}{{\mathbf n}}
\newcommand{\xx}{{\mathbf x}}
\newcommand{\ff}{{\mathbf f}}
\newcommand{\dd}{\text{d}}
\newcommand{\bmnabla}{{\bm \nabla}}
\newcommand{\YY}{{\mathbf Y}}

\newcommand{\cA}{{\mathcal A}}

\newcommand{\cAt}{{\mathcal A}_t}

\newcommand{\ot}{\Omega_t}

\newcommand{\oo}{\Omega_0}

\newcommand{\nx}[1]{{\bm \nabla}_{\xx}\cdot #1}

\newcommand{\nabx}[1]{{\bm \nabla}_\xx #1}

\newcommand{\ddtY}[1]{\frac{\partial #1}{\partial t}{\biggr \vert}_{\YY} }

\newcommand{\Rd}{\mathbb{R}^d}

\newcommand{\uhr}{(\hat{\uu} \circ \cAt^{-1})}

\begin{document}

\begin{frontmatter}
\journal{\it J. Sci. Comput.}
\volume{27}
\issue{1}
\pubyear{2006}
\firstpage{137}
\lastpage{149}
\title{ \bf Mesh update techniques for free-surface flow solvers using spectral element method}

\author[EPFL]{Roland Bouffanais\corauthref{cor}},
\corauth[cor]{Corresponding author.}
\ead{roland.bouffanais@epfl.ch}
\author[EPFL]{Michel O. Deville}
\ead{michel.deville@epfl.ch}
\address[EPFL]{Laboratory of Computational Engineering,\\ \'Ecole Polytechnique F\'ed\'erale de Lausanne,\\ STI -- ISE -- LIN, Station 9,\\ CH--1015 Lausanne, Switzerland}
\begin{abstract}
This paper presents a novel mesh-update technique for unsteady free-surface Newtonian flows using spectral element method and relying on the arbitrary Lagrangian--Eulerian kinematic description for moving the grid. Selected results showing compatibility of this mesh-update technique with spectral element method are given.

\end{abstract}
\end{frontmatter}

\section{Introduction} 
\indent
\indent
Incompressible free-surface flows are encountered in a wide range of engineering and environmental flows. In the nineties the more specific case of turbulent free-surface flows started to be investigated with numerical computation based on high-order methods \cite{LeeWingHoPateraCMAME_1990,HodgesStreetJCP151_1999}. In our work, we aim at computing large-eddy simulation (LES) of unsteady, incompressible and Newtonian turbulent free-surface flows by using the spectral element method (SEM) \cite{PateraJCP54_1984,MadayPateraASME_1989}. The choice of interface-tracking technique was made to ensure an accurate description of the free surface.

This paper highlights the computational techniques we are developing for simulating incompressible free-surface flows using the SEM. These techniques include the arbitrary Lagrangian--Eulerian (ALE) formulation
\cite{Donea_1983,HirtJCP14_1974,RamaswamyKawaharaFEF7_1987}, mesh update and re-meshing methods \cite{GulerBehrTezduyarCM23_1999,JohnsonTezduyarCMAME119_1994}.

This paper is organized as follows. The governing equations in the ALE framework for general free-surface flows are introduced in Section \ref{GoverningEquations}. Then, we present the discretization methods and numerical technique in Section \ref{section3}. Sections \ref{MGT} and \ref{MTO} are dedicated to the moving-grid technique and the mesh-transfer operation, respectively.
\section{Governing Equations} \label{GoverningEquations}
\indent
\indent
A moving boundary-fitted grid technique has been chosen to simulate the free surface in our computations. This choice of a {\it surface-tracking} technique is primarily based on accuracy requirements. With this group of techniques, the grid is configured to conform to the shape of the interface, and thus adapts continually (at each time step) to it and therefore provides an accurate description of the free surface to express the related kinematic and dynamic boundary conditions.

The free-surface incompressible Newtonian flows that we have considered are governed by the Navier--Stokes equations comprising the momentum equation and the divergence-free condition. In the arbitrary Lag-~rangian--Eulerian (ALE), a mixed kinematic description is employed: Lagrangian description of the free surface $\partial \Omega_F(t)$, Eulerian description of the fixed domain boundaries $\partial \Omega_D$ and mixed description of the internal fluid domain $\Omega(t)$, subset of $\Rd$ with $d=2,3$ the space dimension, $t$ referring to the time as the fluid domain is changing when its boundaries are moving. Let us denote by $\oo$ a reference configuration (for instance the domain configuration at initial time $t=t_0$). The system evolution is studied in the time interval $I=[t_0,T]$. The position of a point in the current fluid domain $\Omega(t)$ is denoted by $\xx$ (Eulerian coordinate) and in the reference frame $\oo$ by $\YY$ (ALE coordinate). Let $\cA$ be a family of mappings, which at each $t\in I$ associates a point $\YY \in \oo$ to a point $\xx \in \ot$:
\begin{equation} 
\cAt\ : \oo \subset \Rd \rightarrow \ot \subset \Rd, \qquad \xx(\YY,t)=\cAt (\YY).
\end{equation} 
$\cAt$ is assumed to be continuous and invertible on $\overline{\Omega}_0$ and differentiable almost everywhere in $I$. The inverse of the mapping $\cAt$ is also continuous on $\overline{\Omega}_0$. With these notations the set of equation reads:
\begin{align}
\ddtY{\vv} + (\vv-\ww ) \cdot \nabx{\vv} &=  - \nabx{p} + 2 \nu \nx{\textbf{D}_\xx (\vv)} + {\mathbf f}& & \text{in}\quad \Omega (t),\label{NS} \\ 
\nx{\vv} &=  0& & \text{in} \quad \Omega (t), \label{DF}
\end{align}
with $\vv(\xx,t)$ the velocity field, $p(\xx,t)$ the pressure field (normalized by the constant density $\rho$), $\mathbf D_\xx(\vv)=\frac{1}{2} ( \nabx{\vv} + \nabx{\vv}^T)$ the rate-of-deformation tensor, $\nu$ the kinematic viscosity of the fluid and $\mathbf f$ the body force. The ALE mesh velocity $\ww(\xx,t)$ appearing in \eqref{NS} is defined as
\begin{equation}
  \label{eq:w}
  \ww (\xx, t) = \ddtY{\xx} = \ddtY{\cAt}.
\end{equation}
Surface tension effects are assumed to be negligible as we deal with turbulent flows.
The associated boundary conditions are:
\begin{itemize}
\item[$-$] the kinematic boundary condition on $\partial \Omega_F(t)$: 
\begin{equation}\label{KBC}
\vv \cdot \nn=\ww \cdot \nn,
\end{equation}
$\nn$ being the local outward unit normal to the free surface;
\item[$-$] the dynamic boundary condition on $\partial \Omega_F(t)$: 
\begin{equation}\label{DBC}
-p\nn +2\nu {\mathbf D_\xx}(\vv) \cdot \nn=\textbf{0},
\end{equation}
assuming an inviscid air and zero ambient pressure;
\item[$-$] homogeneous Dirichlet boundary condition on $\partial \Omega_D$:
\begin{equation}\label{HDBCVW}
\vv=\ww={\mathbf 0}.
\end{equation}
\end{itemize}
In addition to the set of governing equations \eqref{NS}--\eqref{HDBCVW}, the closure of this free-surface problem based on a moving-grid formulation requires one more equation governing the evolution of the mesh velocity $\ww$ in the internal fluid domain $\Omega (t)$. The boundary values of $\ww$ being prescribed by the equations \eqref{KBC} and \eqref{HDBCVW} on the boundary $\partial \Omega_F(t)\cup \partial \Omega_D$ of the fluid domain. This last governing equation for $\ww$ will be presented in detail in Section \ref{MGT}.

As our focus is on transient problems, proper initial conditions at time $t=t_0$ for the fluid velocity $\vv$ and for the mesh velocity $\ww$ have to be provided. The initial fluid velocity must satisfy the divergence-free condition and the values of the initial mesh velocity have to be given together with the initial shape of the free surface.

Based on the strong formulation of this free-surface problem given above, one can derive the more appropriate weak transient ALE formulation:\\
\textit{Find $(\vv(t),p(t))\in H^1_{0,D}(\Omega (t))^d \times L^2 (\Omega(t))$ such that for almost every $t\geq t_0$}
\begin{align} \label{WNS}
\frac{\dd}{\dd t} \int_{\Omega (t)} \uhr \cdot \vv \, \dd \Omega\  +\int_{\Omega (t)}\uhr \cdot \nx{[\vv\vv - \vv \ww]} \, \dd \Omega &=  \nonumber  \\
\int_{\Omega (t)} ( p \nx{\uhr} -2 \nu {\mathbf D}_\xx(\uhr) : \nabx{\vv} )\, \dd \Omega&\\
+\int_{\Omega (t)} \ff \cdot \uhr\, \dd \Omega \ \ \ \ \ \ \ \ \ \ \ \ \ \ \ \ \ \ \ \ \ \ \ \forall \hat{\uu} \in&\, H_{0,D}^1(\oo)^d,\nonumber
\intertext{and}
-\int_{\Omega (t)} q \nx{\vv} \, \dd \Omega = 0 \ \ \ \ \ \ \ \ \forall q \in &\, L^2 (\Omega (t)).\label{WDF}
\end{align}
with the functional space $H_{0,D}^1(\Omega (t))$ defined by
\begin{equation*}
H^1_{0,D}(\Omega(t)) = \{v \in L^2 (\Omega (t)), \ \ \nabx{v}\in L^2 (\Omega (t))^d,\ \  v_{{\vert}_{\partial \Omega_D}}=0 \}.
\end{equation*}
It is worth noting that the weak formulation \eqref{WNS}--\eqref{WDF} is only valid in our particular case where homogeneous natural and essential boundary conditions, respectively \eqref{DBC} and \eqref{HDBCVW} are applied to the system. 

\section{Numerical technique and discretization}\label{section3}
\indent
\indent
A classical Galerkin approximation is applied to the set of governing equations in its weak transient ALE form \eqref{WNS}--\eqref{WDF} on the flow domain $\Omega (t)$, in order to determine the pressure and the fluid velocity, keeping in mind that the mesh velocity is obtained by the moving-grid technique developed in the next section. The Galerkin approximation is then discretized by using the spectral element method with the classical staggered ${\mathbb P}_N - {\mathbb P}_{N-2}$ approach to avoid the development of spurious pressure modes. Discontinuous and continuous approximations are respectively taken for the pressure and fluid velocity. The mesh velocity is discretized using the same polynomial space as the fluid velocity, namely ${\mathbb P}_N$, based on a Gauss--Lobatto--Legendre (GLL) grid of order $N$. For the discontinuous approximation of the pressure, a Gauss--Legendre (GL) grid of order $N-2$ is used. Consequently the ALE Navier--Stokes semi-discrete equations can be derived from \eqref{WNS}--\eqref{WDF}:
\begin{equation} \label{SDNS}
\frac{\dd}{\dd t} ({\mathbf M}\underline{\vv})  +{\mathbf C}(\underline{\vv},\underline{\ww})\underline{\vv}= - {\mathbf K} \underline{\vv} + {\mathbf D}^T\underline{p} + \underline{\mathbf F},
\end{equation}
\begin{equation} \label{SDDF}
- {\mathbf D}\underline{\vv} = {\mathbf 0},
\end{equation}
$\mathbf M$ denoting the mass matrix, $\mathbf K$ the direct stiffness matrix, ${\mathbf D}^T$ the discrete gradient operator, $\mathbf D$ the discrete divergence operator, ${\mathbf C}(\underline{\vv},\underline{\ww})$ the discrete convective operator depending both on the fluid and mesh velocities and $\underline{\mathbf F}$ the discrete body force. The update of the position $\xx$ of the mesh points is performed by integrating the following discrete equation:
\begin{equation} \label{SDP}
\frac{\dd \underline{\xx}}{\dd t} = \underline{\ww}. 
\end{equation}

The set of semi-discrete equations \eqref{SDNS}--\eqref{SDP} is discretized in time using a decoupled approach: the linear Stokes computation (linear viscous diffusive term) is integrated based on an implicit backward differentiation formula of order 2, the nonlinear convective term is integrated based on a simple method used by Karniadakis et al. \cite{KIOJCP97_1991}, consisting in an explicit extrapolation of order 2. Finally the update of the position of mesh points is based on an explicit and conditionally stable Adams--Bashforth of order 3 (AB3).

Lastly the treatment of the pressure relies on a generalized block LU decomposition, using a standard fractional-step method with pressure correction.
\section{Moving-grid technique} \label{MGT}
\indent
\indent
As already mentioned in the previous sections, our free-surface flow computations are of interface-tracking type and rely on a moving-grid technique, allowing large amplitude motions of the free surface, generating a grid conforming to the shape of the free surface for an accurate and easy application of the boundary conditions on $\partial \Omega_F(t)$. Moreover a description as accurate as possible of the turbulent free-surface boundary layer is essential to our work. These points  justify by themselves the choice of a moving-grid technique that increases the difficulty of the marginally intractable problem of turbulent viscous flow computations.

The computation of the mesh velocity $\ww$ in the internal fluid domain $\Omega (t)$ is the corner-stone of the moving-grid technique developed in the framework of the ALE formulation. The values of the mesh velocity being prescribed on the boundary $\partial \Omega (t) = \partial \Omega_F (t) \cup \partial \Omega_D$ as expressed by equations \eqref{KBC} and \eqref{HDBCVW}, the evaluation of $\ww$ in $\Omega (t)$ can be obtained as the solution of an elliptic equation:
\begin{equation} \label{ELLIPTIC}
{\mathcal E}_\xx\ww = {\mathbf 0} \qquad \text{in} \quad \Omega (t).
\end{equation} 
This elliptic equation constitutes a classical choice for calculating the mesh velocity \cite{LeeWingHoPateraCMAME_1990}. In the present case it is desirable to impose an additional constraint to the mesh velocity problem, in order to ensure the incompressibility of the mesh by imposing a divergence-free condition to $\ww$:
\begin{equation}
\nx{\ww} =0\qquad \text{in}\quad  \Omega (t).
\end{equation}
Our choice for the elliptic operator $\mathcal E$ is based on the assumption that the motion of the mesh nodes is equivalent to a steady Stokes flow, corresponding physically to an incompressible and elastic motion of the mesh. The boundary-value steady Stokes problem for the mesh velocity can be formulated as follows:
\begin{align}
\ww \cdot \nn &= \vv \cdot \nn & & \text{on }  \partial \Omega_F(t), \label{SETFIRST}\\
\ww \cdot {\bm \tau} & = 0 & & \text{on } \partial \Omega_F(t),\\
\ww & ={\mathbf 0} & & \text{on } \partial \Omega_D,\label{HDBCWV}
\intertext{where $\bm \tau$ is the local unit vector directly orthogonal to $\nn$, and}
\nx{\tilde{\bm \sigma}} &= {\mathbf 0} & & \text{in } \Omega (t), \\
\nx{\ww} &= 0 & & \text{in } \Omega (t), \label{SETLAST}
\end{align}
denoting by $\tilde{\bm \sigma}$ the Cauchy stress tensor of the mesh defined by:
\begin{equation}
\tilde{\bm \sigma} = -\tilde{p} {\mathbf I} + \tilde{\nu} ( \nabx{\ww} + \nabx{\ww}^T)
\end{equation}
with $\tilde{p}$ and $\tilde{\nu}$ being respectively the fictitious mesh pressure and the fictitious kinematic viscosity of the mesh, characterizing the elasticity of the mesh in its motion.

The choice of this boundary-value problem for the mesh velocity has several justifications. Constraining the elliptic equation by a divergence-free condition for $\ww$ allows to ensure the conservation of the volume of the spectral elements, condition that is helpful in practice to have rapidly convergent computations \cite{DevilleFischerMund}. In general the global volume of the computational domain may not be conserved, e.g. with an inflow-outflow imbalance, which requires \eqref{HDBCWV} to be relaxed. In addition, the mesh velocity $\ww$ appears in the convective part of equations \eqref{NS}, \eqref{WNS} and \eqref{SDNS}, together with the divergence-free fluid velocity $\vv$. Moreover it is worth remembering that the divergence-free condition imposed to $\ww$ leads to a conservation of the metrics (the Jacobian being constant in time) when moving the mesh. Finally the unavoidable issue of fulfilling the geometric conservation law (GCL) in the ALE framework \cite{FarhatGeuzaineGrandmontJCP174_2001,FarhatGeuzaineCMAME193_2004,FormaggiaNobileCMAME193_2004} is automatically solved when considering a divergence-free mesh velocity as a consequence of the work of Formaggia and Nobile in \cite{FormaggiaNobileCMAME193_2004}.

From a numerical point of view, the problem corresponding to the set of equations \eqref{SETFIRST}--\eqref{SETLAST} is discretized using the SEM, with a staggered grid ${\mathbb P}_N - {\mathbb P}_{N-2}$ for the couple mesh ($\ww$, $\tilde{p}$). An Uzawa decoupling technique is employed for the treatment of the fictitious pressure.

\begin{figure}[htbp]
  \begin{center}
    \input{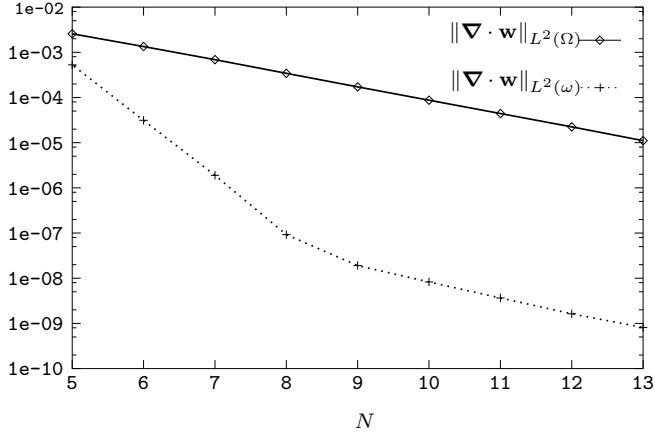}
    \caption{$L^2$-norms of the divergence of the mesh velocity $\ww$ versus polynomial interpolation order $N$ (Log scale)}
    \label{DivergenceW}
  \end{center}
\end{figure}

Based on the technique described earlier, we have developed the following moving-grid algorithm:
\begin{enumerate}
\item Input data: mesh ${\mathcal M}^n$ at $t=t_n$, with nodal coordinates $\xx^n$, fluid velocity $\vv^n$ on $\partial \Omega_F^n$, mesh velocity $\ww^n$ in $\Omega^n \cup \partial \Omega^n$;
\item Step 1: steady Stokes computation of $\ww^{n+1}$ by Eqs. \eqref{SETFIRST}--\eqref{SETLAST};
\item Step 2: update of the nodal coordinates Eq. \eqref{SDP}; spectral element vertices are moved according to the AB3 scheme:
\begin{equation}
\xx^{n+1} = \xx^n + \frac{\Delta t}{12} (23 \ww^{n}-16\ww^{n-1}+ 5 \ww^{n-2}) ;
\end{equation}
\item Creation of the new mesh ${\mathcal M}^{n+1}$ with the new Gauss--Lobatto and Gauss--Lobatto--Legendre grids for each new spectral element;
\item Output data:  mesh ${\mathcal M}^{n+1}$ at time-step $t=t_{n+1}$, with nodes coordinates $\xx^{n+1}$, mesh velocity $\ww^{n+1}$ in $\Omega^{n+1} \cup \partial \Omega^{n+1}$.
\end{enumerate}
Two performance tests have been carried out on a study case where one edge of a squared mesh is deformed by a sine profile. Both of these tests aimed at verifying the spectral element volume conservation that is theoretically imposed by the divergence-free condition on $\ww$. The first test is dedicated to the verification of the global volume conservation, by computing the relative change of the volume of the computational domain when moving the grid from the initial square to the deformed one. For several number of spectral elements and for a polynomial interpolation order ranging from 1 up to 12, the relative change of the volume of the computational domain is found to be smaller then the machine precision. The second test is also devoted to the volume conservation but now from a local perspective and by numerically computing the $L^2(\Omega)$- and $L^2(\omega)$-norm of the divergence of the mesh velocity $\ww$ for a polynomial interpolation order $N$ ranging from 5 up to 12, where $\omega$ is interior of the computational domain made of the spectral of elements of $\Omega$ not sharing an edge with $\partial \Omega$. Results are presented on Fig. \ref{DivergenceW} and it is found that these norms are exponentially decreasing with $N$ as expected when using a spectral element method \cite{DevilleFischerMund}. Moreover we can note that the $L^2(\omega )$-norm of $\bmnabla \cdot \ww$ has a faster rate of convergence than the $L^2(\Omega )$-norm. This is justified by the fact that the divergence-free constraint cannot easily be enforced at the grid points located in the vicinity of the boundaries of the computational domain $\Omega$.
\vspace{0.3cm}
\section{Mesh-transfer operation} \label{MTO}
\indent
\indent
In the previous section was presented the moving-grid technique used in our work to move the grid points at each time-step, generating a new mesh. Depending on the amplitude of the mesh deformation at each time-step, this technique can be applied during an important number of iterations. Nevertheless the mesh obtained by moving the grid nodes can be too convoluted therefore affecting the accuracy and the convergence of the simulation. Consequently a re-meshing operation is to be called by a specific control parameter (e.g. a discrete Jacobian positiveness criterion) to provide a new mesh topology. Before starting the ALE Navier--Stokes computation at the next time-step on this newly created mesh, it is mandatory to transfer some information from the previous mesh to the new one. The main requirement imposed to this so-called mesh-transfer operation is to conserve the spectral accuracy of the SEM. The information to be transferred comprises six fields: the fluid velocities $\vv^n$, $\vv^{n-1}$ and the mesh velocities $\ww^n$, $\ww^{n-1}$, $\ww^{n-2}$ (time-integration schemes are of order 2 for $\vv$ and 3 for $\ww$) and also the pressure at the current time-step (use of a pressure correction technique). As written in Section \ref{section3}, the velocities are expanded over a GLL grid and the pressure over a GL one. Therefore our mesh-transfer technique must be capable of transferring fields defined over GL and GLL grids.

Our mesh-transfer algorithm for GL grids being based on the one for GLL grids, we will start presenting in detail the latter. Let us consider two meshes ${\mathcal M}^1$ and ${\mathcal M}^2$ corresponding to different mesh topology of the same computational domain and the mesh-transfer operation from ${\mathcal M}^1$ to ${\mathcal M}^2$. In the sequel we will assume that we have the following decompositions in terms of spectral elements:
\begin{equation}
\Omega_i \cup \partial \Omega_i = \overset{E_i}{\underset{e=1}{\bigcup}}\Omega^{i,e} \qquad \text{for } i=1,2.
\end{equation}
As the computational domain remains unchanged, for each spectral element $\Omega^{2,e}$ of ${\mathcal M}^2$ we have:
\begin{equation}\label{inclusion}
\Omega^{2,e} \subset (\Omega_1 \cup \partial \Omega_1) \qquad \forall e=1,\dots , E_2.
\end{equation}
Due to Eq. \eqref{inclusion} our mesh-transfer technique only requires an interpolation procedure. Let us note the physical location of the set of GLL grid points of a spectral element $\Omega^{2,e_2}$ ($e_2=1,.\cdots, E_2$) by $\{\xx_{ij}^{2,e_2} \}$ with $(i=1, \cdots, N_{x,2}+1; j=1, \cdots, N_{y,2}+1)$, $N_{x,2}$ (resp. $N_{y,2}$) being the order of the polynomial interpolation in the $x$-direction (resp. $y$-direction) for the mesh ${\mathcal M}^2$ (with the same notations, $N_{x,2}$ and $N_{y,2}$ can be different from $N_{x,1}$ and $N_{y,1}$ respectively). The proposed algorithm can be summarized in three steps:
\begin{enumerate}
\item Find the spectral element $\Omega^{1,e_1}$ of ${\mathcal M}^1$ containing $\xx_{ij}^{2,e_2}$;
\item Determine the position $\rr^{1,e_1}$ of $\xx_{ij}^{2,e_2}$ within the parent element $\hat{\Omega}^{1,e_1}$ of ${\Omega}^{1,e_1}$;
\item Compute the value of the field at the point $\xx_{ij}^{2,e_2}$ given $\rr^{1,e_1}$, the GLL Lagrangian interpolation basis and the values of the field at the GLL grid points of $\Omega^{1,e_1}$.
\end{enumerate}
The first step causes no difficulty in its implementation. The second step uses a transfinite interpolation procedure in each spectral element, in order to invert the iso-parametric mapping $\bm \Phi$:
\begin{equation}
\rr^{1,e_1} = (r^{1,e-1},s^{1,e_1})= {\bm \Phi}^{-1} (\xx^{2,e_2})\quad \text{with} \quad \rr^{1,e_1}\in \hat{\Omega}^{1,e_1} = [-1,1]^2.
\end{equation}
In practice, the inversion is carried out differently depending on the topology of the spectral element. With quadrangular spectral elements, our algorithm performs a direct analytical inversion of the affine mapping $\bm \Phi$ which is computationally inexpensive. With deformed spectral elements \cite{DevilleFischerMund}, the inversion of $\bm \Phi$ relies on the so-called `inverse iso-parametric mapping technique' from Lee and Bathe \cite{LeeBathe1994} which is based on a Newton--Raphson type iterative procedure.

Finally in the last step, efficient routines compute the following spectral interpolation:
\begin{equation}
\uu(\PhiPhi^{-1}(\xx_{ij}^{2,e_2}) ) = \uu ( r^{1,e_1}, s^{1,e_1} ) = \sum_{k=0}^{N_{1,x}}  \sum_{l=0}^{N_{1,y}} \uu_{kl} \, \pi_k(r^{1,e_1}) \pi_l(s^{1,e_1}),
\end{equation} 
with $\{ \pi_j (\xi)\}_{j=0}^{N_{p,1}}$ and $p=x,y$, the one-dimensional GLL Lagrangian interpolation basis of degree $N_{p,1}$. As said earlier the mesh-transfer technique for GL grids relies on the one for the GLL grids. In our simulations, the only GL-interpolated field that has to be mesh-transferred is the pressure field. Therefore, by interpolating the pressure on the GLL grid, then by applying the GLL mesh-transfer operation introduced earlier and finally by interpolating back on the GL grid, we manage to perform the requested operation. It is important to minimize the occurrence of a re-meshing as our mesh-transfer technique is computationally expensive even for quadrangular elements (affine iso-parametric mapping). A more detailed assessment of the performance of this technique is provided at the end of this section.

This mesh-transfer operation has been extensively tested in order to ensure its compatibility with the SEM, regarding its exponential rate of convergence. Tests involving the following two key parameters have been carried out: the polynomial interpolation order $N$ and the amplitude of the change in topology of the grid when re-meshing.

\begin{figure}[htbp]
    \input{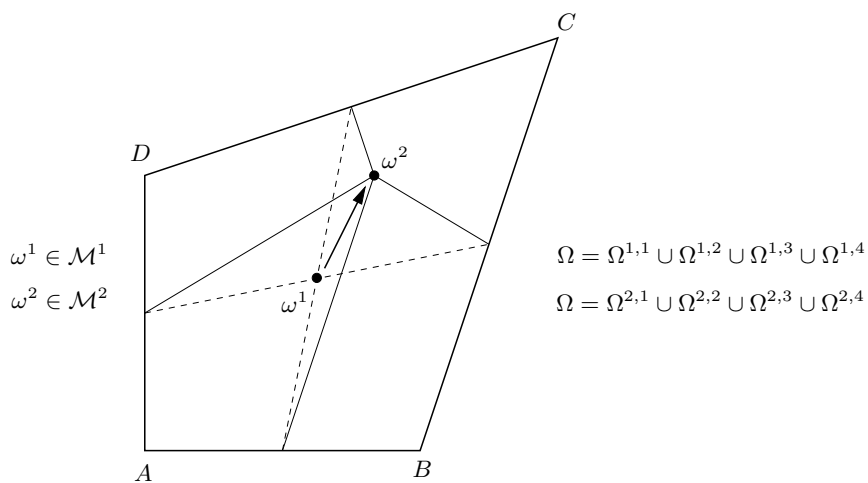}
    \caption{Sketch of the computational domain $\Omega$, the two meshes ${\mathcal M}^1$ and ${\mathcal M}^2$ and their spectral element decompositions before and after a prescribed re-meshing operation obtained by moving the central vertex $\omega$}
    \label{oldnewmesh}
\end{figure}

The set-up is presented in Fig. \ref{oldnewmesh} and is made of a mesh comprising four spectral elements. The change in topology of the mesh is prescribed by moving only the vertex $\omega$ (see Fig. \ref{oldnewmesh}) common to all four spectral elements and afterwards the mesh-transfer operation is performed. 
\vspace{0.5cm}

To evaluate the dependence of our technique with the interpolation order $N$, the central vertex is moved to produce a topological change in the mesh by a factor of approximately 10 \%. An analytical field $f$ is interpolated on the initial mesh and mesh-transferred onto the distorted mesh, leading to the interpolated field $\tilde{f}$. The interpolation error is defined by $\varepsilon=\| f - \tilde{f} \|_{L^2(\Omega )}$ and computed values are presented in Table \ref{errororder}, showing a conservation of the exponential rate of convergence.

\vspace{0.5cm}
\centerline{\begin{tabular}[h]{cccc} 
    \hline\hline
    \mbox{} & & & \\[-2ex]
    $N$ & $\varepsilon = \| f - \tilde{f} \|_{L^2(\Omega )}$  & $N$ & $\varepsilon = \| f - \tilde{f} \|_{L^2(\Omega )}$ \\[1ex]
    \hline\hline
    \tt 3 & \tt 7.232e-03  & \tt 12 & \tt 4.400e-12 \\
    \tt 4 & \tt 1.487e-03  & \tt 13 & \tt 9.850e-14 \\
    \tt 5 & \tt 1.367e-04  & \tt 14 & \tt 1.252e-14 \\
    \tt 6 & \tt 2.307e-05  & \tt 15 & \tt 3.602e-15 \\
    \tt 7 & \tt 1.457e-06  & \tt 16 & \tt 3.354e-15 \\
    \tt 8 & \tt 2.067e-07  & \tt 17 & \tt 1.843e-15 \\
    \tt 9 & \tt 8.382e-09  & \tt 18 & \tt 1.585e-15 \\
    \tt 10 & \tt 1.172e-09  & \tt 19 & \tt 1.105e-15 \\
    \tt 11 & \tt 3.383e-11  & \tt 20 & \tt 1.151e-15 \\ 
    \hline\hline
  \end{tabular}}
\begin{table}[h]
  \caption{\label{errororder}Evolution of the error $\varepsilon$ with the spectral interpolation order $N$}
\end{table}

To characterize the effect of the distortion of the mesh on our mesh-transfer operation, all possible positions of the moving vertex within the computational domain $\Omega$ were envisaged. In particular, we present here the case where $\omega$ is moved along the diagonal $AC$ of the computational domain $\Omega$ as shown in Fig. \ref{oldnewmesh}. Its motion is characterized by the set of coordinates $(\alpha, \beta)$ of $\omega$ in the parent domain $\hat{\Omega}=[-1,1]^2$. The interpolation error $\varepsilon$ was again computed for three values of $N$ and results appearing in Table \ref{errormotion}, show that our technique is totally independent on the amplitude of topological change of the mesh due to the re-meshing operation.

\vspace{0.5cm}
\centerline{\begin{tabular}[h]{cccc} 
    \hline\hline
    \mbox{} & & & \\[-2ex]
    $\alpha=\beta $ 
    & \multicolumn{3}{c}{${\varepsilon = \| f^2 - \tilde{f}^2 \|}_{L^2(\Omega )}$} \\
    \cline{2-4}
    & $N=8$ & $N=10$ & $N=12$ \\[1ex]
    \hline\hline
    \tt -0.9 & \tt 2.100e-07& \tt 1.005e-09 & \tt 3.849e-12\\
    \tt -0.8 & \tt 2.267e-07 & \tt 1.111e-09 & \tt 4.448e-12\\
    \tt -0.7 & \tt 2.000e-07 & \tt 1.006e-09 & \tt 3.906e-12\\
    \tt -0.6 & \tt 1.928e-07 & \tt 1.106e-09 & \tt 4.073e-12\\
    \tt -0.5 & \tt 2.289e-07 & \tt 1.033e-09 & \tt 3.786e-12\\
    \tt -0.4 & \tt 1.847e-07 & \tt 1.053e-09 & \tt 4.199e-12\\
    \tt -0.3 & \tt 2.326e-07 & \tt 1.160e-09 & \tt 4.166e-12\\
    \tt -0.2 & \tt 2.231e-07 & \tt 1.204e-09 & \tt 4.332e-12\\
    \tt -0.1 & \tt 2.067e-07 & \tt 1.172e-09 & \tt 4.400e-12\\ 
    \tt 0.0 & \tt 4.563e-16 & \tt 1.199e-15 & \tt 7.886e-16\\
    \tt 0.1 & \tt 2.067e-07 & \tt 1.172e-09 & \tt 4.400e-12\\
    \tt 0.2 & \tt 2.231e-07 & \tt 1.204e-09 & \tt 4.332e-12\\
    \tt 0.3 & \tt 2.326e-07 & \tt 1.160e-09 & \tt 4.166e-12\\
    \tt 0.4 & \tt 1.847e-07 & \tt 1.053e-09 & \tt 4.199e-12\\
    \tt 0.5 & \tt 2.289e-07 & \tt 1.033e-09 & \tt 3.786e-12\\
    \tt 0.6 & \tt 1.928e-07 & \tt 1.106e-09 & \tt 4.073e-12\\
    \tt 0.7 & \tt 2.000e-07 & \tt 1.006e-09 & \tt 3.906e-12\\
    \tt 0.8 & \tt 2.267e-07 & \tt 1.111e-09 & \tt 4.448e-12\\
    \tt 0.9 & \tt 2.100e-07 & \tt 1.005e-09 & \tt 3.849e-12\\
    \hline\hline
  \end{tabular}}
\begin{table}[h]
  \caption{\label{errormotion}Evolution of the error $\varepsilon$ when $\omega$ moves along the diagonal $AC$, with $\gamma=1$ and for three different values of $N$}
\end{table}

Lastly, the computational expense of the mesh-transfer has been evaluated for a polynomial degree $N=10$ in both directions ($19^2$ grid points for this 2D grid), and as previously for a topological change in the mesh by a factor of approximately 10 \%, corresponding to a `small' 2D case. The results confirm the afore-mentioned cost: a complete mesh-transfer corresponds to approximately 100 Navier--Stokes solves depending on the value of the time-step.
\section{Conclusion and future studies} \label{conclusion}
\indent
\indent
A novel isochoric moving-grid technique and mesh-transfer technique for spectral element grids have been presented. Both of these techniques are the corner-stones of our computations of turbulent free-surface flows using spectral element method. Part of the work was to ensure that these two techniques have no effect on the exponential rate of convergence, the main reason of our choice of the spectral element method. We have obtained positive results all along the extensive series of tests carried out to verify the behaviour of this rate of convergence. The development of an automatized re-meshing scheme coupled to a re-meshing control parameter are still under investigations.

Our next goal is to simulate three-dimensional turbulent free-surface flows using the techniques presented in this paper with the difficult task of gaining a better insight into the physics involved in the thin turbulent boundary layer near the free surface.
\section{Acknowledgments}
\indent
\indent
The authors would like to thank Professor Einar R\o nquist and Professor Yvon Maday for fruitful discussions.

This research is being partially funded by a Swiss National Science Fundation Grant (No. 200020--101707), whose support is gratefully acknowledged.

\end{document}